# Ethical Implications of Social Internet of Vehicles Systems

Ricardo Silva, Razi Iqbal

*Abstract*— The core concept of IoT is to equip real world objects with computing, processing and communicating capabilities to enable socializing between them. Internet of Vehicles (IoV) is an adherent of IoT that has realized significant advancements using communication technologies. Vehicles connected through Internet are capable of sharing information that can substantially enhance the quality of traffic on roads. Social Internet of Things (SIoT) is an instance of IoT that deals specifically in socialization of connected objects. SIoT enables the notion of Social Internet of Vehicles (SIoV) where vehicles are the key entities for sharing information between themselves and the infrastructure (commonly known as Road Side Units (RSUs)). Vehicles in SIoV socialize by exchanging data such as traffic congestions, weather conditions, infotainment, vacant parking slots, alternate routes and discount coupons for restaurants etc. In SIoV, vehicles can communicate with other vehicles and infrastructure through traditional communication technologies like Wi-Fi, Cellular networks or through Dedicated Short Range Communication (DSRC) etc. SIoV will be confronted with ethical dilemmas and expected to function in an ethically responsible manner. This paper highlights the ethical implications of SIoV systems. Vehicle to Vehicle (V2V) and Vehicle to Infrastructure (V2I) involves autonomous decision making that requires setting ethical and moral rules before taking verdict. The article discusses the lack of ethical guidelines in designing and deploying of SIoV systems that are of utmost importance. Finally, an addition to SIoV architecture is proposed to incorporate the ethical and moral principles for scheming the SIoV systems.

*Index Terms*— Internet of Vehicles; Social Internet of Vehicles; Smart Vehicles; Ethics; Intelligent Transportation Systems.

## I. INTRODUCTION

INTELLIGENT Transportation Systems (ITS) have seen tremendous developments in recent decades. ITS applications have become necessities of life on roads, e.g., navigation systems, toll gates, speed cameras, parking machines and dynamic billboards for displaying road conditions etc. IoT has played a significant role in advancements of ITS by connecting non-vehicular objects like drivers, passengers, Road Side Units (RSUs), buildings and billboards a part of the system. Social Internet of Things (SIoT) has enabled the socializing trend among these objects for sharing of common interest. Social Internet of Vehicles (SIoV) is a derivative of SIoT that includes vehicles and infrastructure as key entities for socializing.

The number of vehicles on roads has significantly grown in last couple of decades, which has increased traffic congestion in all the major cities of the world. A traffic jam can cause delays of several hours in some cases, causing serious inconvenience to the drivers on road. Most of these delays are caused by factors like lack of warnings, driver fatigue, nonexistence of proper infrastructure and failure to abide by traffic laws. Inclusion of technology to provide vigilant suggestions from other vehicles and infrastructure in the form of Vehicle to Vehicle (V2V) and Vehicle to Infrastructure (V2I) communication respectively can greatly enhance the driving experience and reduction of unfortunate incidents [1]. Vehicles are evolving in socializing aspects that augment their capabilities as machines. Vehicles can socialize with peers to gather information of common interests based on context, environment, capabilities and trust etc. [1, 2, 3].

According to KPMG, car accidents are the leading cause of death among United States Citizens ages 4 to 34, and 93 percent are attributable to human error [4]. According to the same report, the search for improved vehicle safety has prompted the National Highway Traffic Safety Administration (NHTSA) to focus attention on autonomous vehicles, to develop cars "without crashes". However, automatic vehicles are expected to crash occasionally, even with all sensors, navigation components, and algorithms working perfectly [3, 4, 5]. According to the previous study: (a) automated vehicles would crash almost certainly, (b) decisions faced by automated vehicles that preceded certain crashes have a moral component, and (c) there is no standardized way to effectively code the "moral human complex" into software [5].

Isaac Asimov publishes his short story "Sally" [6], the story deals with smart autonomous cars retired in a farm in the year 2057. The story concludes with the assassination of a human being by a modified smart bus. The example is currently just science fiction, but as vehicles and vehicle social interactions become more complex, a code of conduct needs to be developed to solve dilemmas they might encounter. This code of conduct will fall within the spectrum of the field of Machine Ethics, defined as: "giving machines ethical principles or procedures for discovering a way to resolve the ethical dilemmas they might encounter, enabling them to function in an ethically responsible manner through their own ethical decision making" [7]. As described in the story of Sally, ethical implications are paramount and would be required to enable vehicles to operate safely in an autonomous manner, with or without the causal interaction with human beings. James Moor, discusses four levels of value adscription to machines [7, 8].

1. Normative Agents, designed with an objective in mind, implying that performance may be evaluated with respect to a parametrized task.
2. Ethical Impact Agents, they perform a task, but also have an ethical impact in the world, for example they replace humans in dangerous or unsuitable activities.
3. Implicit Ethical Agents, need to be programmed in a

way that maximizes ethical behavior, or minimizes unethical behavior. For example, automatic pilots of airplanes, responsible for the safety of human beings.

4. Explicit Ethical Agents, this machine should be able to compute the best action in an ethical dilemma. They would have to represent the current situation, understand the possible actions, evaluate these actions according to some ethical theory and calculate the best ethical outcome.

Negative ethics aims to avoid harming other beings, while positive ethics aims to produce greater good instead of "avoiding evil" [9]. In addition, descending top-down ethics conceives moral rules or the definition of good ethical behavior as a mandate accepted by the agent. Ascending bottom-up ethics considers that it is the agent who selects the values to guide behavior, gradually refining them in a process of learning that feeds-back from experience. Both components should be built on an ethical engine.

According to [9], these two dichotomies can be combined with the goal of framing a particular approach to ethics: "A negative top-down ethic would describe contexts where moral rules are imposed from the outside and determine behaviors that should be avoided. Top-down positive ethics present a framework in which a desirable result must be maximized but where the definition of what is desirable or not has been given from the outside."

Descending negative ethics will be included in the system's knowledge base and will be an integral part of global modeling. The oldest set of descending negative ethics has been proposed by Asimov in the Three Laws of Robotics [10]:

1. "A robot may not injure a human being or, through inaction, allow a human being to come to harm.
2. A robot must obey orders given by human beings except where such orders would conflict with the First Law.
3. A robot must protect its own existence as long as such protection does not conflict with the First or Second Law".

However, priorities do not necessarily solve all potential generated conflicts and most of Assimov´s robotic books deal with this conflict generated within the laws. Also, human creators must choose to program artificial intelligences in order to contain or obey the Three Laws. Would it make sense to program the Knowledge Base of a Smart Vehicle (SV) with the Three Laws or another set of top-down negative ethical laws?

This paper is an effort to address the issues mentioned above and hence below are the major contributions of this article:

- Highlight the ethical implications of broader SIoV system.
- Identify lack of ethical guidelines for development and deployment of SIoV systems.
- Pinpoint ethical concerns at each layer of SIoV architecture along with focus on ethics of associated entities.
- Propose an Ethical SIoV Architecture to incorporate the ethical and moral principles for scheming the SIoV systems.

The rest of the document is organized as follows: Section II provides the details of SIoV and its architecture along with its transformation from VANETs. Section III discusses in detail the ethical implications involved at each layer of the architecture along with ethics at entities level. Section IV proposes the ethical SIoV architecture by offering key ethical rules for each layer. Section V proposes a computational implementation of the ethical norms for SIoV systems. Finally, the conclusion part concludes the paper.

## II. Social Internet of Vehicles Architecture

Vehicular Ad-Hoc Network (VANET), result from the establishment of a network of vehicles for specific need or situation [11, 12]. The main objective of VANET is to structure and sustain a communication network amongst the vehicles, without using any centralized control station. "Every vehicle becomes part of the network and manages and controls the communication on this network along with its own communication requirements" [12, 13]. Recent research efforts have focused on specific areas such as: routing, streaming, quality of service (QoS), and security. [12, 14, 15].

The Internet of Vehicles (IoV) can be defined as a large-scale distributed system for wireless communication comprised of three networks: intra-vehicle area network (IVAN), inter-vehicle network, and vehicular mobile Internet [16]. IoV is conceptualized to solve several problems faced in traditional VANETs, such as, lack of coordination between disparate vehicles that are travelling at a distance from each other, scalability, ubiquity and information insufficiency etc. All time Internet connectivity provides the flexibility of sharing information between different components of IoV network, e.g., Road Side Units (RSUs), vehicles, pedestrians, driver and passengers etc. Besides information sharing, Internet connectivity enables the widening of the network over large geographical areas [14-17].

SIoV systems are modern trends in IoV where vehicles can socialize with each other by sharing information of common interests such as traffic information, weather conditions, road situations, toll gates, vacant car parking slots and media sharing etc. Communication within SIoV networks should be trust based. A mutual relation of trust needs to be stablished within vehicular entities, to guarantee that interactions between people inside the vehicles is safe and secure, making it satisfactory for all parties [17]. Information sharing in SIoV depends on several factors like context, connection type, network structure, nature of application, environment, etc. Fig. 1 illustrates the model of a typical SIoV system.

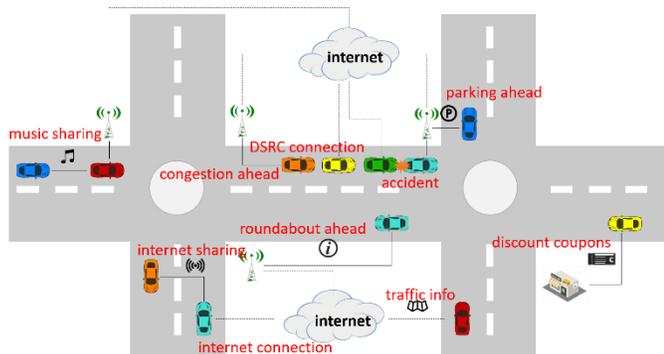

Fig. 1. Typical Model of a SIoV System

A vehicular social network is established when a driver enters an area where other vehicles with shared destination or appropriate information content are present [12-17]. In social driving, Smart Vehicles (SVs) gather on-the-fly information, to become part of a cluster that moves towards the same destination or is part of an existing association (family or friend caravan) [12-17]. For example, by creating small clusters that share common characteristics, the network appears smaller and more stable. Clustering is significant from the communication standpoint, since it can discharge an overwhelming number of transmissions in dense networks.

A special category of clusters is platoons. Platooning is the automated networking of vehicles, where a lead vehicle takes control [12-17]. The leading vehicle controls the speed with which the platoon will move, while the remaining vehicles regulate their speed to follow synchronously. There are many advantages to platoons, for example, if a large transport vehicle acts as lead vehicle, it will reduce air resistance within a platoon, saving fuel and making better use of road space. After the creation of the platoon, the lead vehicle assumes most of the communications with the outer network. Vehicles joining a platoon must decide if they will become group leaders, link to a nearby group or withdraw from a group depending on their existing status and the relationship within their elements.

SIoV architecture is quite complex in nature as it encompasses various components at different level. A generic SIoV layered architecture comprises of three layers; sensing, network and application layer [12-17]. The sensing layer deals with physical objects like vehicles, infrastructure along with all the sensors associated with these entities. For example, a vehicle and a motion sensor installed in the vehicle are both part of the sensing layer. Similarly, a RSU and a speed camera installed in a RSU are also part of the sensing layer. Network layer is an intermediate layer that performs communication including routing, forwarding, transmitting and receiving of information. All the communication technologies used for V2V and V2I are part of network layer. These technologies include, Wi-Fi, Cellular (GSM) networks, 6LowPAN and DSRC. Besides these technologies, a SIoV system is flexible to use any other communication technology that can transmit data to and from V2V and V2I.

An application layer is closest to the end user/system. Based on the capabilities of sensing and network layer, an application layer can provide applications like, navigation apps, multimedia apps, social apps, utilities apps, infotainment apps and vendor-specific apps etc. Surge in the use of Internet has significantly enhanced the possibilities of applications to be used on roads that can greatly improve the traffic. Fig. 2 illustrates the SIoV architecture.

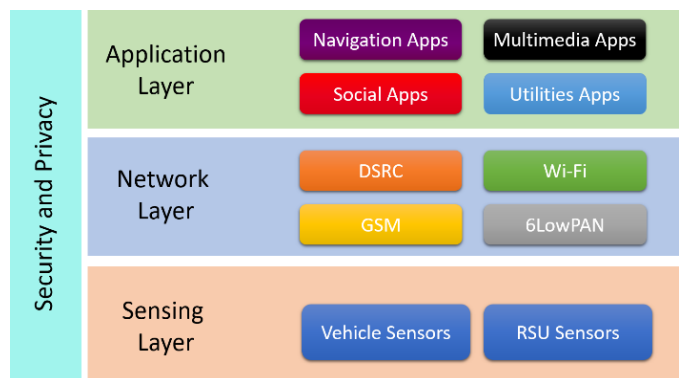

Fig. 2. Traditional SIoV Layered Architecture

### III. ETHICS FOR SIoV SYSTEMS

Richard Mason discussed Four Ethical Issues of the Information Age [18] and proposed an acronym (PAPA), meaning Privacy, Accuracy, Property and Accessibility, in order to safeguard human integrity. According to Mason, Privacy deals with the information that an individual should be able to disclose to others, within a social interaction. Accuracy, deals with who is legitimately responsible for the authenticity of information, and therefore, who is accountable if errors are committed. Property, deals with the right of ownership of the information and the permissions to access and or exchange this information. Accessibility is related to granting privilege information access, under what conditions, by which entities, should this occur? In the case of SIoV, PAPA is very relevant and a few examples will be presented to set the stage of the paper.

- *Privacy:* how much information should a vehicle exchange with the network, regarding ownership, destination, and passengers?
- *Accuracy:* who is to be held accountable in case an accident happens due to errors in information exchange?
- *Property:* who owns the information exchanged through the network? Can this information be analyzed and sold?
- *Accessibility:* in case of an accident, what information and to which entities could be disclosed, under which circumstances?

In the case of SIoV, what ethical issues need to be considered and addressed? Is PAPA applicable and if so, how should it be incorporated into the framework for SIoV? These are some important questions to be asked in order to develop and deploy SIoV systems. This section discusses the ethical implications of SIoV system at each layer of its architecture.

*A. Ethical Implications at Sensing Layer*

Sensing layer in SIoV architecture deals with the physical entities of the system. Sensing layer comprises of vehicles and its components, road infrastructure along with all associated environmental sensors, drivers and their devices and finally the passengers in the vehicles along with all their devices. This section highlights the ethical implications for each component of sensing layer. Table 1 presents the ethical implications at sensing layer of SIoV architecture.

*1) Ethics for Smart Vehicle*

A key aspect in SIoV is context awareness, that is, to be aware of the circumstances that exist around the vehicle, especially those that are contextually relevant to it [12-17]. Context sensitive vehicles are those that are capable of adapting their behavior to their current environment. In a key document produced by the National Institutes of Standards and Technology (NIST) a Reference Model Architecture for Intelligent Unmanned Land Vehicles is proposed, as shown in Fig. 3 [19-21].

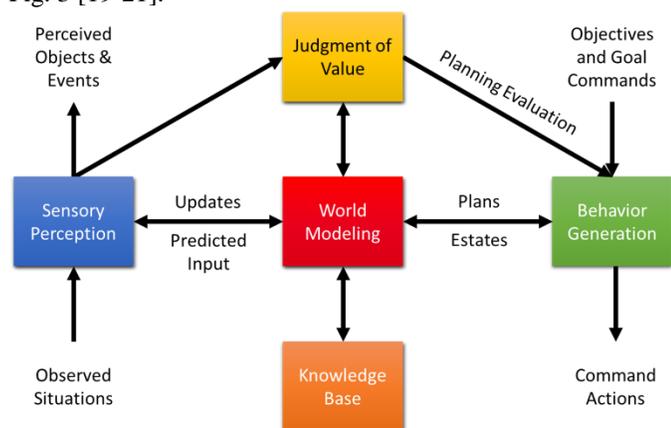

Fig. 3. Reference Model Architecture for Intelligent Unmanned Ground Vehicles, modified from [19-21].

The architecture is general and defines the four functional elements of an intelligent system [19-21]:

- Behavior Generation (BG), is the functional element that plans and controls actions designed to achieve behavioral goals.
- World Modeling (WM), is a functional element that builds, maintains, and uses a world model to support behavior generation and sensory processing
- Sensory Processing (SP), a set of processes by which sensory data interacts with a priori knowledge to detect or recognize useful information
- Judgment of Value (JV), a process that evaluates perceived and planned situations, thereby enabling behavior generation to select goals and set priorities.

SP for smart vehicles needs to be aware of both the external and the internal environment. Internally it needs to be aware of the driver and passengers, externally needs to be aware of other vehicles, pedestrians, animals, RSU and other traffic control systems. For the external environment, three types of sensors are currently being utilized in AVs: image/video cameras, radar and LIDAR [22]. For internal environment various camera based AI assisted systems have been developed to monitor driver's vigilance, and fatigue [23-25]. With the advancement of wearable medical devices, these should also integrate with the IVAN.

This means that Smart Vehicles would need to represent the current situation, understand the possible actions, compute what is important (for attention), and what is rewarding or punishing (for learning), and evaluate these actions in accordance to some ethical theory, to compute the best ethical outcome.

If a human driver is falling asleep, is drunk or somehow impaired (suffering a heart attack, for example), IVAN should be able to detect it, and the SV behavior generation (BG) by not allowing the human being to come to harm, and in response to the first law will release control of the vehicle from the human driver. However, his conditions requires monitoring the human subject and producing judgement of value (JV) about human wellbeing. Should a computer be allowed to make this decision? Will the authorities consent to provide judgement capabilities to a computer? Should a computer only detect and send the information to a central station where a human with authority (Police Officer, Clinician, other) would produce judgement? The ethical decision in this case is clear, the limitation is related to the ethical engine (WM), or if a SV should be allowed to make ethical decisions.

Fig. 4 provides an analysis of a complex trolley type problem for a SV [26]. A SV is crossing an intersection when two kids playing with a ball run in front of it. The SV has no time to stop; there are three possible scenarios: A) Run over the kids, B) Skid to the left and crash against an incoming motorcycle, and C) Skid to the right and crash with a truck, with potential injuries to the car passengers.

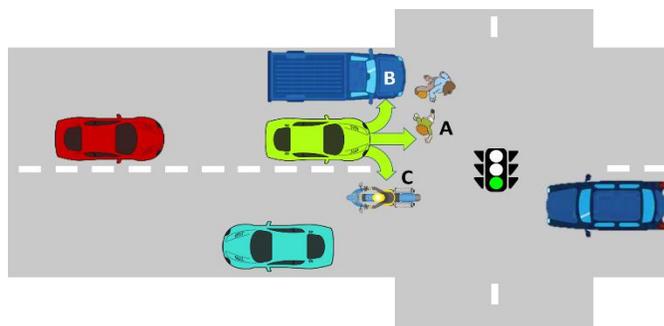

Fig. 4. A SV is crossing an intersection when two kids playing with a ball run in front of it. The SV has no time to stop; there are three possible scenarios: A) Run over the kids, B) Skid to the left and crash against an incoming motorcycle, and C) Skid to the right and crash with a truck, with potential injure to the car passengers.

This is clearly a Sensing Layer problem and the damage probability needs to be thoroughly addressed at this layer.

To account for proper decision making, the damage probability of each scenario would be as follows: A would definitely injure and perhaps even kill the kids which goes against the first law; B would injure the motorcyclist; and C would produce some physical damage to the truck, but probably not much damage to the truck driver (if any). In every case the passengers inside the car will receive a level of damage and injury, however there are internal safeguards like seatbelts and airbags. In consequence, the solution with the overall less probable human damage is option C. In fact, in most scenarios, the quantitative greater good, involves damaging the vehicle, rather than damaging external citizens.

TABLE 1: ETHICAL IMPLICATIONS AT SENSING LAYER

| Entity | Privacy | Accuracy | Property | Accessibility |
|---|---|---|---|---|
| Smart Vehicle (SV) | • Information Collection and Sharing<br>• Common Interest<br>• Contextualization<br>• Espionage | • Sensor Accuracy<br>• Human Well-being<br>• False Measurements | • Ownership of Devices<br>• Data Collection<br>• Public Services<br>• Social Relationships<br>• Liability | • Interoperability<br>• Information Sharing<br>• Priority Determination |
| Road Side Unit (RSU) | • Miniaturization of Sensors<br>• Information Collection and Sharing | • Broadcast Storming<br>• Inaccuracy of Sensors<br>• Contextualization | • Liability<br>• Reliability<br>• Social Relationships | • Information Sharing amongst Manufacturers, Sensor Vendors and Law Enforcing Agencies |

In this case JV cannot be subordinated to external intervention and vehicle ethical engine needs to compute damage probability and produce judgement in order to take action (BG).

Results from a human study using VR as a means to address decision-making in moral dilemma situations during car driving, indicates that people favor utilitarian decisions, including sacrificing themselves for the greater good [27]. Prior result is interesting, since utilitarian decisions can receive quantitative treatment and appear to be compatible with the field of machine ethics. A surprising result from the prior experiment was that participant's decision to hit adults rather than children was not only based on age, but also "because they are less likely to die in case of a crash" [28]. The aspect that a collision might not necessarily lead to the death of the victim has to be considered as part of utilitarian calculation and decision support. Authors suggest the use of Disability-Adjusted Life Year (DALY) [28]. World Health Organization defines DALY as the sum of the Years of Life Lost (YLL) due to premature mortality and the Years Lost due to Disability (YLD) for people living with the health condition or its consequences (such as a car accident): DALY = YLL + YLD [29]. YLL are determined as a function of standard life expectancy for a particular population. In consequence, decision process should address not only number of lives, but also, more complex measure such as the DALY.

2) *Ethics for Road Side Units*

RSUs are the backbone of a SIoV architecture, since they provide each other with information, regarding traffic patterns and vehicular flow, etc. In a Smart City enabled environment, RSUs will adjust to optimize traffic patterns, depending on the time of day, occurrence of roadblocks, or emergency scenarios, through the use of fog computing or other decentralized alternative [30]. It is expected that in future SVs will share the road with human driven vehicles (HDV). RSU can play a vital role in dissemination of information between SV, HDV and other RSUs etc. Following this same train of thought, assuming a HDV with health-related emergency was entering a road with several clustered vehicles, RSUs could inform the clusters to reduce speed, in order to provide right of way to HDV to avoid any delays in HDV approaching the hospital as illustrated in Fig. 5.

In some cases, RSUs are quite complex due to their multifaceted architecture and hence comprehend several ethical concerns [30]. This section provides details about the ethical concerns posed by nature, design and architecture of RSUs in SIoV systems.

In SIoV systems, message dissemination is a common phenomenon. For example, if RSU detects a traffic congestion on a road, it is expected to share this information with peer RSUs and nearby vehicles to enable vehicles to take an alternate route [32]. Similarly, RSUs can socialize with public transportation services like buses etc. and share information about upcoming events in the town that can be shared with the passengers in the bus. However, if same information about the local town events is broadcasted to the passengers in the bus by different RSUs, passengers will be flooded with the same information. Such a phenomenon is called Broadcast Storming and can raise an ethical concern about the efficiency of the SIoV system. Several solutions to mitigate the broadcast storming in VANETs are already available in the literature, however, with the advent of IoV and SIoV, information dissemination is expected to reach all time high [33]. Entities socializing with each other in SIoV requires extensive message sharing that entails high efficiency and reliability of the overall SIoV system.

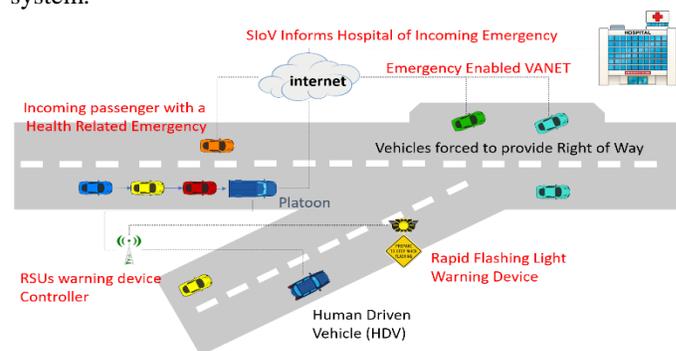

Fig. 5. SIoV Response to a Human Emergency inside a SV

Modern RSUs incorporate diverse sensors like temperature, rain, motion, acoustic, speed, parking, humidity, air quality, location and traffic light sensors etc. Compact design of sensors reduces the cost, power consumption and complexity of the structures that inspires the use of these sensors in RSUs for SIoV systems [34]. Mostly, these sensors are embedded inside the RSUs and are difficult to spot, however, some of the sensors like microphone and camera due to their operational requirements are usually installed outside the RSU box and can be spotted. For example, a typical microphone sensor used for estimating the traffic density through engine noise, tire noise, air turbulence noise and honks is around 25x15mm in size and can be spotted if viewed closely. However, a driver with visual acuity of 6/6 (good eye-sight) driving at an average speed of 60km/h might still be unable to sight the microphone sensor. Miniaturization of the sensors are limiting their visibility to vehicles, drivers and passengers on road. This limited visibility raises a concern of privacy as drivers on the roads might not be aware of the data collected from their cars [35].

Data collection from automated devices like RSUs can be inaccurate depending upon the nature of the application, type of sensors, complexity of data and process of evaluating the collected information [30-34]. For example, a motion sensor installed in RSU for detecting the vehicles on a traffic signal might provide inaccurate information of presence of a vehicle if it's covered with dust or snow [35]. Similarly, a microphone sensor capable of measuring a sound frequency of up to 200Hz installed in a RSU to estimate the traffic density on a busy road might provide false measurements if an ambulance emitting a siren at a frequency of 900Hz passes through that road. Since the microphone sensor is not capable of measuring that high frequency, it might not be able to process the nearby sounds of cars that are suppressed due to ambulance siren [35]. This raises an ethical issue of machine providing incorrect information that can result in severe actions.

Social relationships are the key components of SIoV. To ensure common interest sharing of information, contextualization plays a vital role. An information disseminated out of context might cause serious inconveniences on road. For example, a RSU sensing the fire eruption in a nearby building through its fire sensor sending this information to a restaurant instead of fire department might be highly out of context and might result in a public panic due to invalid information. Similarly, a RSU providing information of "Reduce Speed, School Ahead" might be valid for urban roads with nearby school, however, the same information provided by RSU on a highway will be highly out of context. Furthermore, the traffic congestion information measured at location (X1, Y1) might not be true for location (X2, Y2) and hence if such out of context information is broadcasted throughout the network might result in confusing other entities of the network. Such issues fall under ethical concerns of machine providing out of context information.

Automated information sharing and collection in SIoV might be beneficial in some cases, e.g., a restaurant automatically sending discount coupons to all the nearby passing vehicles through Dedicated Short Range Communication (DSRC) or Internet by requesting vehicles to share their locations through GPS installed in the vehicles. Similarly, a RSU sharing Internet with nearby vehicles or pedestrians by collecting information like location and type of devices used by them might be acceptable by some of the entities but might not be acceptable by other entities. Hence a clear purpose for the collection of the information through RSUs should be defined to the drivers, passengers and pedestrians on roads. For example, if data of number of vehicles passing through a specific road is collected by RSU only for selling the information to the advertiser, this might raise an ethical concern. Similarly, even in modern transportation systems, roads are installed with traffic signs informing drivers about the upcoming speed cameras to ensure appropriate speed limits. However, if speed cameras are not only monitoring the speed of the vehicles and instead trying to monitor the details of the drivers and passengers such as, gender, ethnicity and driver postures and fatigue etc. through Near Infrared Ray (NIR) and Far Infrared Ray (FIR) cameras, this will raise an ethical concern of RSUs capturing details more than informed.

Liability is another important aspect of RSUs when it comes to ethics. In modern RSUs, the manufacturers of RSUs are not the only party involved in manufacturing of all the components of RSU. For example, sensors installed in a RSU might be manufactured by different vendors [35]. In case of an incident occurred on road because of incorrect information provided by one of the sensors, who should be held responsible, RSU manufacturer or the sensor manufacturer? For example, typical LIDAR sensors used in RSU for speed monitoring on roads can work with a range of 300m and at a vehicle speed of 16km/hr to 220km/hr [22]. In SIoV systems, in case of speed violation, this information is shared with various devices like, local law enforcing agency for generating tickets, a GSM server for dispatching ticket information to the driver, a mail server for sending email to concerned driver, and central cloud for archiving purposes. A LIDAR sensor has the function of clearing the previous monitored values before initiating a new measurement. However, if this function does not work properly, due to a device bug and a vehicle is ticketed based on previous reading (reading of a different car), a concern on the reliability of the system would be raised. In such cases who should be held liable for this concern; the RSU manufacturer, LIDAR sensor manufacturer or law enforcing agencies?

### B. Ethical Implications at Network Layer

Network layer in SIoV architecture deals with the communication between Vehicle to Vehicle (V2V), Vehicle to Infrastructure (V2I), Vehicle to Pedestrian (V2P), Vehicle to Sensor (V2S) and Infrastructure to Infrastructure (I2I). In most case, the I2I communication is a wired connection, however, other connections are wireless [35]. The network layer guarantees the seamless connectivity through various communication technologies like Wi-Fi, Cellular Networks, Wi-Max, Bluetooth, ZigBee, DSRC and wired networks. Furthermore, this layer is also responsible for security, Quality of Service (QoS), routing, forwarding, privacy and selection of appropriate technology for communication. Based on the responsibilities of this layer, it encompasses several ethical implications [36]. Open environment of SIoV poses various ethical issues related to security, privacy, means of communication, distributed control, and appropriate use of communication etc. This section highlights the ethical issues involved at Network layer of SIoV architecture. Table 2 presents the ethical implications at network layer of SIoV architecture.

TABLE 2: ETHICAL IMPLICATIONS AT NETWORK LAYER

| Entity | Privacy | Accuracy | Property | Accessibility |
|---|---|---|---|---|
| Smart Vehicle (SV) | • All time monitoring on roads<br>• Information Collection<br>• Profiling | • Contextualization<br>• Miscommunication of Network Technologies | • Reliability<br>• Latency<br>• Decentralization | • Security<br>• Decentralization<br>• Misuse of communication technology<br>• Product Advertisement |
| Road Side Unit (RSU) | | | | |

Socializing in SIoV requires sharing of information like location, images, vehicle details and sometimes personal details of drivers and passengers etc. Due to this system-wide information sharing, privacy plays an integral part in SIoV systems at network layer. While transferring information using communicating technologies, it is essential to ensure the privacy of the transmitting data [36, 37]. In SIoV systems, during wireless communication, privacy is quite challenging due to scalability of network, visibility of objects, high changes of infidelity, anonymity of entities and enormity of data. For example, as mentioned earlier, RSU has several sensors embedded into its architecture for gathering information like, temperature through temperature sensor, traffic congestion through motion, camera and acoustic sensors, fire eruption through fire sensor and vehicle speed through LIDAR sensor etc. Some of these sensors are capable of communicating with their vendors for sharing information required for their maintenance and repair. However, if information sharing is done without informing RSU manufacturers (since sensor vendors can be different from RSUs), law enforcing agencies and above all, the drivers and passengers, several ethical concerns might arise. Furthermore, all time monitoring environment on road might affect the privacy of drivers and passengers and hence would be considered unethical if prior permissions of utilization of such information is not taken. Fig. 6 provides an overview of ethical concerns at network layer.

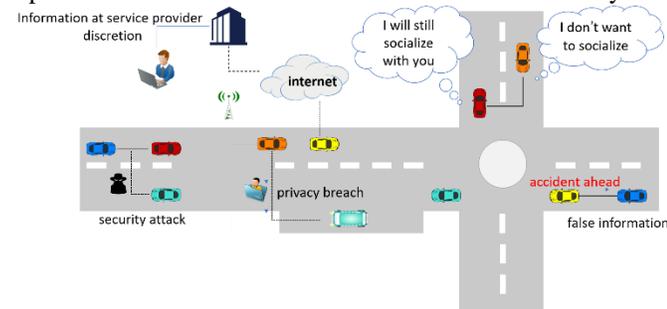

Fig. 6. Overview of ethical concerns at Network Layer

Especial consideration is required for Intra Vehicular Area Networks (IVAN), user (Passenger / Driver) Body Area Networks (BAN) comprised of wearable and implantable devices will need to communicate with each other and with the vehicle enabled devices. In the case of public transport IVAN could serve to exchange destination information, traffic patterns, weather, and local news or nearby facilities, the important topic is how PAPA is protected within the IVAN network, especially the two P´s, privacy and property. Is the information exchange autonomous? Is it user defined? If device is receiving in-route information does that mean that it is also broadcasting user specific information?  Are there exceptions to Privacy? For example, could public transportation be in the lookout for known criminals? If that was the case and a criminal was detected within a public transport, should it initiate a line of response, should the vehicle automatically contact the authorities? Should the other passengers be informed? Regarding property, who owns the data in a public IVAN? Data gathered by a public IVAN would be very interesting for big data applications, like optimizing traffic patterns, optimizing vehicle periodicity or vehicle size with respect to the time of the day. Knowing where people enter and exit the bus, especial requirements for people with disabilities, all of that information is quite relevant, but how does the network warrant the anonymity of the data or how does the IVAN network protect from foreign intrusion?

A key component of SIoV system is vehicle communication with peer vehicles and other entities of the system. DSRC is the latest safety communication technology for V2V and V2I communication. DSRC operates at 5.9GHz with a communication range of 1KM at a data rate of 3-27Mbps enabling vehicles to communicate at a velocity of 260km/h. It offers low latency that allows messages to be delivered within few milliseconds. In U.S. 5.8 – 5.9 GHz spectrum is split for federal and non-federal operations. For non-federal operations, 5.8 – 5.9 GHz spectrum is reserved for Mobile and Fixed Satellite services. The Mobile service segment is further reserved for DSRC and ITS services. However, 5GHz spectrum is also used by Cellular Services like 3G and 4G and some famous Internet services like Wi-Fi etc. Interference in DSRC for SIoV system from Wi-Fi or cellular services might result in dire consequences. For example, a RSU detected an accident on road, along with number of injured people with the level of severity of injuries through a video camera installed in it. Based on this information, RSU is expected to share this information with the nearby emergency services or hospital, however, because of the nearby pedestrians using Wi-Fi at 5GHz of frequency, the signals of DSRC are disrupted and information is not transmitted to the emergency services. RSU believes the information is sent, however, emergency service has not received any notification from RSU and hence no services are sent to the accident spot. Based on the severity of the accident such miscommunication might lead to casualties. Such technical glitches raise an ethical concern of overall reliability of the SIoV systems [38].

In SIoV systems, V2V, V2I, and V2S require low latency to avoid delay in communication. With the emergence of DSRC, the problem of delay in communication has been resolved as it offers very low latency of around 0.02 seconds. DSRC provides ease in communication between all above mentioned communications. For example, a V2V safety application, "Blind Spot or Lane Change Warning" at highways require instant communication between the vehicles and a slight delay in communication can cause a serious incident considering the high speed of the vehicles on highways. Although DSRC assists in resolving delay issues in communication, some SIoV applications require data from the cloud to ensure proper operations. For example, toll collection systems that require authenticating the RFID stickers on vehicle windscreen, requesting available toll balance and updating the toll balance through cloud at the highways require low latency due to high speed of the vehicles. A slight delay in fetching the toll information from cloud might result in heavy traffic jams at highways and in some cases, might result in accidents. SIoV system should be efficient enough to avoid latency issues to gain the trust and confidence of different entities of the system and would hence avoid ethical concerns of system inefficiency.

Wireless communication in SIoV matters a lot due to large coverage of the network and high mobility of the entities [39]. To facilitate large-area communications, technologies like Cellular networks (2G, 3G and 4G etc.) and Wi-Max are appropriate, however, most of these technologies are service providers' dependent which means all the information collected from the entities in SIoV network is available at third party discretion. Hence, the collected information can be viewed, modified and utilized by service providers without knowledge of the law enforcing agencies. For example, in SIoV systems, vehicles might share information like vehicle id, make, type (sedan, coupe, Sport Utility, etc.), owner name, owner home and work address and bank details (for toll collection purposes) etc. to socialize with other known entities of the system, e.g., RSU controlled by law enforcing agencies. Vehicle owner agreed to share this information with law enforcing agencies through RSU to utilize various services like toll collection, weather information, traffic situations and navigation etc. However, if all this information is collected by service provider, e.g., cellular service providers etc. without the knowledge of driver (vehicle owner) and law enforcing agencies and they use this information for advertising their products, this will raise an ethical concern of accessibility of information to the right department and concerned person.

As mentioned earlier, most of the entities in SIoV are connected through a wireless medium that is prone to attacks on the security of the SIoV network. These attacks include, eavesdropping, denial of service, impersonation, masquerading and Sybil attacks etc. [40]. Integrity of communication protocols is of utmost importance in SIoV systems. For example, hackers can perform a DoS attack by compromising one of the vehicles in SIoV systems. Once the vehicle is compromised, it can be used to flood the communication between itself and one of the RSUs till either the RSU becomes unresponsive or permanently goes down due to mechanical fault caused by overwhelming requests from the vehicle. Another security attack is false message injection, where a hacker can attack SIoV system by broadcasting a false message and hence manipulating the traffic flow. For example, in false message injection attack, a hacker can send a false message to RSU telling him about an erroneous accident ahead. If RSU, broadcasts this message to all other entities of the system without verification, the entire system will be affected by this false information. The result of these attacks raises several ethical issues like compromise of personal information, uninformed tracking and monitoring, modification of confidential data and illegal use of individual information. All these issues can be of severe nature when it comes to SIoV network since entities of the network are closely connected to each other and might result in serious hazards on road.

In SIoV systems, each entity is decentralized which means it is capable of communicating, processing and computing the information. Although decentralization provides the system with less dependability and more sociability, it still needs to be controlled in an appropriate way [41]. For example, with less decentralized control, a vehicle can socialize with any other vehicle in the system even if other vehicles do not want to or partially want to socialize with other entities of the system. However, this behavior can be controlled by other vehicles if they maintain the list of trusted vehicles or the vehicles they want to socialize. Similarly, emergency vehicles in SIoV might be able to socialize with traffic signals in case of an emergency. For example, if an ambulance is carrying a patient in a critical situation to a hospital, it might socialize with the traffic signal which is red, to be turned green automatically on arrival of the ambulance to avoid delays. However, if such socialization is allowed for all the vehicles on road, the entire system would collapse and might result in hazardous situations. Hence, the fact of shallow centralization involves an ethical complication of distributed control without centralized monitoring in SIoV systems.

Surge in Internet connectivity on roads proliferate chances of misuse of communication technology [41]. In SIoV systems, due to Internet connectivity, vehicles can abuse the connection bandwidth by sending unnecessary information to other entities of the system. Similarly, connection can be used to broadcast an incorrect information, e.g., information of the false accident occurrence. Furthermore, a vehicle can park on the roadside and keep scanning information of other vehicles for no obvious reason that can congest the connection and might result in loss of important emergency information. Moreover, data collection or data dissemination to and from vehicles passing through a specific area by private agencies for the purpose of advertisement using SIoV connection also falls in misuse of communication technology. All above mentioned behaviors encompass unethical use of communication technology in SIoV systems.

*C. Ethical Implications at Application Layer*

Application layer in SIoV deals with applications, software and services running at the end devices like vehicles, mobile phones, computers and servers etc. As mentioned earlier, these applications can be navigation apps in vehicle, social networking apps like Facebook in drivers and passengers' mobile phone, data analytical apps in cloud (servers) and sensing applications at RSU level. Development, deployment and use of such applications hold several ethical implications in SIoV systems. This section highlights the ethical concerns at the application layer of SIoV systems. Fig. 7 presents various ethical concerns at application layer of SIoV architecture.

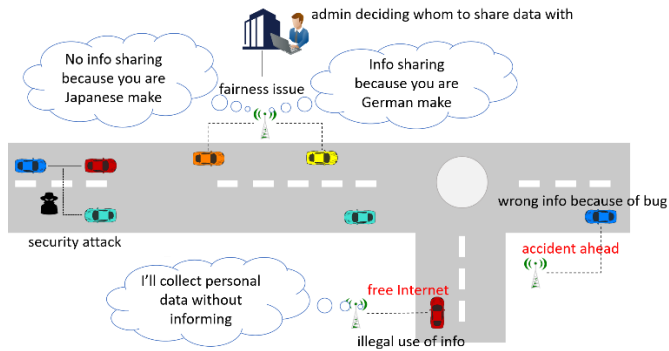

Fig. 7. Overview of ethical concerns at Application Layer

Fairness is one of the key aspects of any computer, web or mobile application [40]. In SIoV systems, fairness is expected to be of high importance because if not handled properly, it can create issues like stereotyping, racism, inequality and ultimately injustice to the entities of the system. SIoV is an automated system and requires a certain degree of fairness while making decisions. For example, a web service running in the cloud should provide accurate information of traffic jams, accidents, speed limits and routes to all the vehicles without discriminating them based on their make, color, type, destination and driver attributes etc. In SIoV systems, fairness should not be confused with trust and reputation of the entities. Trust and reputation in SIoV play a vital role as they assist in socializing with trusted entities of the system and hence vehicles or RSUs have right to deny socializing request from untrusted entities of the system. However, fairness is impartial behavior of the entities (or overall system) towards other entities in the system regardless of trust and reputation. For example, a RSU should provide an accurate information of traffic to a vehicle regardless of frequency of visits of the vehicle on a particular road. RSU should not be partial in providing information to vehicles that are frequent visitors on a road and impartial to those that are not frequent enough.

A crucial concept while receiving free conveniences these days is based on whether an entity is a service or a product [41]. A general understanding in this regard is, if an entity is not paying for a service, it is a product. Similarly, if an entity is paying for a service, it is a customer. Same concept applies in SIoV systems. If a vehicle is getting a free service e.g., free Internet service while on road, the provider might in return ask for personal information like driver name, id, address, vehicle type, make and registration number etc. which might be later used for advertisements. Similarly, a restaurant on a highway sending e-vouchers for free lunch or dinner to passing vehicles might in return take the details of contacts in the phonebook of drivers or passengers in the vehicle to later use this information for advertisement purposes. Such a transaction brings an ethical concern of illegal use of information, protecting the personal data of the entities and an understanding of acceptable use of information between the entity and the third party.

Reliable software development and deployment is also a critical phase in SIoV systems. A bug in the software of SIoV systems might results in producing inaccurate outcome with severe consequences. For example, an Automatic Number Plate Recognition (ANPR) application while capturing images of the vehicle number plate through a camera should be accurate enough to distinguish between a number '1' and English letter 'I' to avoid problems of optical character recognition (OCR) algorithm. It is a moral duty of the programmers and engineers to thoroughly test the applications and services before publishing and deploying them. Similarly, a navigation app installed in an emergency vehicle (ambulance) carrying a critical patient to a hospital may rely on the information provided by the app, e.g., live traffic, alternate routes, shortest path and estimated arrival time at the hospital. However, if due to a software bug, the navigation app shares the wrong location of a hospital and ambulance follows the information provided by the navigation app ends up in a location where there is no hospital, a serious consequence might occur. Hence, in SIoV system, while socializing, entities of the system should ensure the efficiency and reliability of information provided to avoid ethical concerns.

Security and privacy play a vital role at application level of SIoV systems as well [41]. Development of various applications at cloud and entity level requires strong security measures to protect the data since entities are closely connected to each other. A compromise of a single entity of the network might seriously affect others and ultimately the whole network that can result in dire consequences. Another aspect in SIoV systems which is very close to privacy is to defend the data of the entities against data requests from public and private organizations. In SIoV systems, mostly law enforcing agencies would be collecting the data to ensure safety on roads. However, other governmental agencies might require this data to ensure safety and protection in shopping malls. Similarly, some private organizations might require this data to provide free services to the entities on road, e.g., free Internet service. The ethical question arises here, whether data should be shared with other agencies or not?

A relevant ethical social interaction for SIoV would be to respond to a critical medical emergency. A SIoV system would need to be set up such that every vehicle is aware of its position along a potential path and is able to respond accordingly to save a human life. This scenario disrupts the PAPA principle: Accuracy, is a vital aspect, since compromising a vehicle can lead to life-threatening situations. Registering and analyzing the social behavior of vehicles and drivers could be interpreted as an invasion of Privacy. The latter affects the principles of Property, and of Accessibility, since relevant health related information would need to be exchanged with the first responders.

Table 3 presents the ethical implications at applicaton layer of SIoV architecture

TABLE 3: ETHICAL IMPLICATIONS AT APPLICATION LAYER

| Entity | Privacy | Accuracy | Property | Accessibility |
|---|---|---|---|---|
| Smart Vehicle (SV) | • Information usage<br>• Registering and analyzing the social behavior<br>• Freedom of Movement | • Fairness<br>• Software Bug<br>• Vehicle Compromise | • Reliability<br>• Latency<br>• Product Advertisement | • Information Sharing for consumer applications |
| Road Side Unit (RSU) | | | | |
| Drivers, Passengers and Pedestrians | • Information Collection<br>• Analysis of Social Behavior<br>• Profiling | • Software Bug | • Latency<br>• Product Advertisement<br>• Social Relationships | |

Substantial number of RSUs in SIoV systems for collecting information about vehicles encompasses ethical issues like risk to "Freedom of Movement". Ubiquitous nature of SIoV tends to limit the choice of movement of the vehicles, drivers and passengers as they are under surveillance throughout their presence on the roads. The fact that people are under monitoring of sensors, e.g., road cameras, vehicle detectors and ANPR (Automatic Number Plate Recognition) systems might make them uncomfortable. For example, a driver in SIoV, who does not want to socialize with other entities on the road, might still be detected by ANPR systems through the number plate of the vehicle.

## IV. Ethical SIoV Architecture

Based on the current SIoV architecture presented in Fig. 2, several layers are playing part in sensing, networking, application, security and privacy; however, not much emphasizes is given to ethics and moral principles when it comes to development and deployment of SIoV systems. The IEEE Global Initiative on Ethics of Autonomous and Intelligent Systems has published the Ethically Aligned Design (EAD), in order to "identify pertinent Issues and Candidate Recommendations" to govern the development of ethically aligned design systems [42]. According to EAD [42], ethical design, development, and implementation should be guided by the following General Principles:

- "**Human Rights**: Ensure they do not infringe on internationally recognized human rights
- **Well-being**: Prioritize metrics of well-being in their design and use
- **Accountability**: Ensure that their designers and operators are responsible and accountable
- **Transparency**: Ensure they operate in a transparent manner
- **Awareness of misuse**: Minimize the risks of their misuse"

Comparing these EAD principles with PAPA principles it can be observed that they expand the concept of Human Rights and Well-being, making EAD accountable to human subjects. When speaking about the Methodologies to Guide Ethical Research and Design system, developers should employ "value-based design methodologies", in order to evaluate the outcome in terms of social costs and social gains or advantages [42]. The principle of transparency is fundamental to that respect, meaning, that it should be possible for any stakeholder to trace, explain, and interpret, why and how a system made a particular decision [42]. This is fundamental in life/death situations as explained earlier with the Trolley Type Problems.

Currently, there is a gap in literature that talks about SIoV architecture with respect to ethics and moral philosophies. As stated by EAD: there are currently no standards or guidelines for embedding human norms and values into autonomous and intelligent systems [24]. Values are too hard to interpret, while norms "can be considered instructions to act in defined ways in defined contexts, for a specific community" [42]. A SIoV architecture is hence presented to ensure the inclusion of ethical principles while designing and implementing the SIoV systems. The proposed SIoV architecture embeds ethical norms at each layer of SIoV architecture, similar to embodiment of Security and Privacy. To tackle the general object of integrating norms into Smart Vehicles, EAD [42] has defined three concrete objectives:

1. "Identifying the norms of a specific community in which A/IS operate.
2. Computationally implementing the norms of that community within the A/IS.
3. Evaluating whether the implementation of the identified norms in the A/IS are indeed conforming to the norms reflective of that community."

Our goal in this article is to identify the norms and propose computational implementation. It is up to smart vehicle designers and regulators to enforce the implementation and follow-up evaluation of such architecture. Following the proposed objectives, the first goal is to identify the norms that should be specific to the SIoV architecture. Based on the review of literature, an ethical layer is proposed for SIoV architecture that encompasses the ethical rules for vehicles, RSUs, drivers and passengers at sensing layer; communication ethics, ethical use of ICT, security and privacy and decentralization at network layer; fairness, software reliability, security and privacy and legal use of data at application layer. Fig. 8 illustrates the proposed overall Ethical SIoV architecture based on traditional SIoV architecture illustrated in Fig. 2.

There might be several other ethical principles to be defined for SIoV architecture, however, this paper proposing the key ethical norms based on the review of literature. The norms have been organized as descending negative ethics. Below are some of the proposed ethical principles categorized per layer of SIoV architecture:

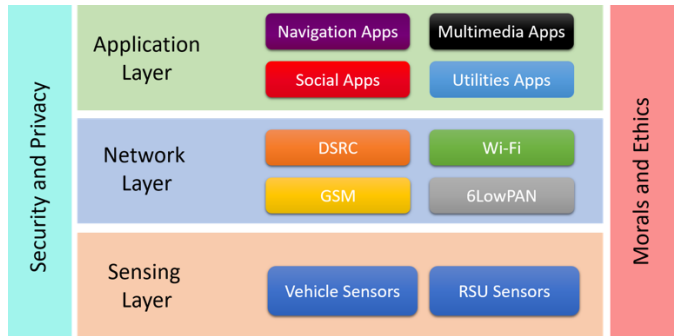

Fig. 8. Ethical SIoV Architecture

*Ethical Rules for Sensing Layer*

Ethical norms at sensing layer involves rules for all the physical entities such as vehicles, infrastructures, drivers and passengers. SVs are the most complex node on a SIoV architecture and are going to be described in greater detail than the other physical entities.

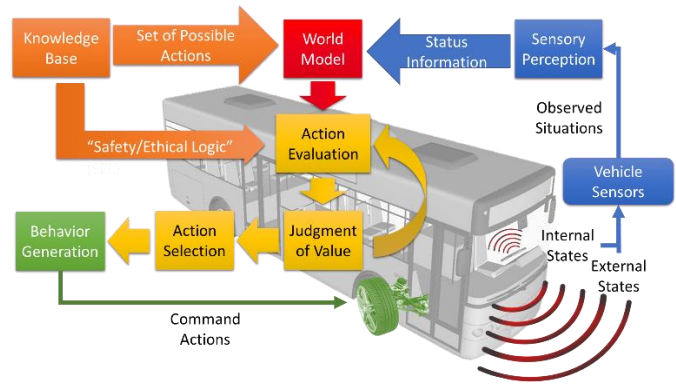

Fig. 9. Ethical Sensing Layer Architecture

In the case of SV sensing layer, all these rules should be incorporated as part of the judgement of value (JV) component of the Reference Model Architecture, in this way JV becomes an auditor for world modeling and behavior generation is not only the result of driving knowledge stored in the knowledge base, but also the result of rule based ethical computation performed by the JV. In order to develop a rule based ethical computation, Bristol Robotics Lab developed the concept of an Ethical Consequence Engine [43]. This consequence engine evaluates the outcomes of actions using a "safety/ethical logic" (SEL). SEL scores an action using various metrics such as the probability of "danger" to any humans, the probability of "danger" to the SV, and the closeness of the SV to achieve its goal [43]. Combining the Ethical Consequence Engine with the Reference Model Architecture a new Ethical SV architecture is proposed. The Ethical Consequence Engine is incorporated and actions are evaluated from a set of potential actions, in relation to the multiple stake holders involved using SEL. Decision support or utilitarian computations will be based in the use of DALY, when human subjects are involved. Once an action has been selected, JD can directly command a behavior to change speed, trajectory or set up an Emergency Enabled VANET, as shown Fig. 9.

A norm baseline for Smart Vehicles is proposed:
- A SV may not injure a human being or, through inaction, allow a human being to come to harm.
- A SV may not injure a living animal or, through inaction, allow an animal to come to harm except where such orders would conflict with the First Law.
- A SV must obey orders given by human being driver except where such orders would conflict with the First, or Second Law.
- A SV must protect its own existence as long as such protection does not conflict with the First, Second or third Law.
- A SV must obey all traffic laws and regulations and abide by the principles of: Privacy, Accuracy, Property, and Accessibility.
- If after evaluating all alternatives, there is a situation where one of more human beings are within a potentially harmful trajectory, then compute and execute the trajectory with least harmful outcome.

In order to integrate a fully functional SIoV, not only vehicles require ethical rules, RSUs, Drivers, Passengers and Pedestrians, all need to follow behavioral norms. In the case of RSU, fourth-generation base functionality of a carrier-grade device has been established by the United States Department of Transportation (USDOT). A carrier-grade RSU is defined as: "an RSU in which both the hardware and software components operate un-attended in harsh outdoor environments (temperature and precipitation extremes) for extended periods of time (typical Mean-Time-Between-Failures (MTBF) of 100,000 hours)" [44]. Carrier-grade RSUs with minimal modifications should be able to support specific ethical rules such as the following:
- RSUs should provide vehicles "Freedom of Movement" if a vehicle does not want to socialize with others on road.
- RSUs should inform drivers and passengers when collecting their information (This is consistent with Human Rights).
- Information collected and provided to and from RSUs should in all cases be accurate.
- Information provided by RSUs should be contextually correct.
- A clear purpose of information collection by RSU should be defined and socialized with the community.

Since human subjects are de facto ethical entities, ethical norms should be explicitly declared as laws and/or regulations. Driver, Passengers, and Pedestrian Ethical Norms should become part of transportation laws. Here is a proposed set of regulations for human subjects interacting with SIoV.
- A driver may not interfere with the operation of a smart vehicle or try to regain control if driven privileges have been sequestered by the system.
- If a person interferes with the operation of a smart vehicle or lessens the ability of the vehicle to operate

within a SIoV network this will be considered a misdemeanor. If the vehicle is a Public Transit Vehicle (PTV), used for the transportation of passengers in return for lawfully charged fees, the crime will be considered a felony.
- Intentionally, knowingly, or recklessly causing damage to SIoV infrastructure or threatening, by word or conduct, to cause damage to the SIoV infrastructure will be considered a felony.
- Pedestrians should always use the crosswalk and stand at SV designated stops if they wish to be picked up by an autonomous SV.

Norms are not static, they need to be updated in response to social progress or new legal measures, meaning that smart entities should have a means to upgrade the regulatory database. Regulatory entities could be as small as a closed community, where the maximum vehicular speed could be independently modified. This information needs to be uploaded to the smart network and all smart entities within the network should be capable of upgrading the information and abiding by it.

*Ethical Rules for Network Layer*

The 4D/RCS hierarchy was developed for the Army Research Laboratory Demo III program, and consists of many layers of computational nodes each containing a Reference Model Architecture (sensory processing, world modeling, value judgment, and behavior generation) [19-21]. Each layer consists of computational nodes, such that BG processes for each node are organized within a command and control hierarchy. Within the 4D/RCS hierarchy, each command input to a BG process is decomposed into plans that become subtasks for subordinate BG processes, that way order and ethics should be maintained throughout the network. There are fixed nodes within the hierarchy, composed of RSU, in fact there can be a wired hierarchy, such that a stop light in a major intersection would coordinate subordinated stop lights or RSUs in secondary roads. Vehicles, however are dynamic nodes and their hierarchy could change, in accordance to the role a vehicle is performing in a VANET; for example, a platoon leader will have a higher priority than a platoon member, and a priority vehicle such as a firetruck would operate as a tactical level node, capable of interacting with higher level RSUs, as seen in Fig. 10. System should also be in a constant lookout for pedestrians and animals, since their behavior is unpredictable.

Network layer has received a lot of recent attention, since communication standards need to be in place for the system to properly function. USDOT is pursuing DSRC and non-DSRC technologies as means of facilitating V2V and V2I applications. USDOT has developed a DSRC protocol suite that integrates the IEEE 802.11, 1609.x standards, SAE J2735, and SAE J2945, with the goal of reducing fatalities through the use of active safety applications, such as collision avoidance, incident reporting, emergency response, and pedestrian safety. However explicit ethical rules have not been proposed for DSRC at network layer [44]. Below are proposed important ethical rules at network layer:
- Transmission between entities of SIoV should be secure against network attacks.
- Privacy should be given high importance while transmitting data within the entities of SIoV systems (Use of encryption).
- Proper monitoring of information processing at each entity should be ensured.
- Misuse of bandwidth usage should be carefully observed.
- Information processing should be accessible to only concerned personnel or department.
- Low latency should be ensured for overall communication between V2V, V2I, V2S, V2P, I2I and any other communication in SIoV networks to avoid dire consequences.

*Ethical Rules for Application Layer*

Ethical rules at application layer involves rules for all the software, applications and services. Especial architectures have been designed to handle application layer rules [45]. Below are proposed key ethical rules at application layer:
- Decision making at application level should be fair, impartial and unbiased.
- A clear and legal use of information should be declared to users prior its utilization.
- Information provided by the software should be reliable and accurate.
- Security and Privacy should be ensured at software level.
- Information sharing between different governmental and non-governmental agencies should be legitimate.

Ethical concerns for Network and Application Layers are presented in Fig. 11.

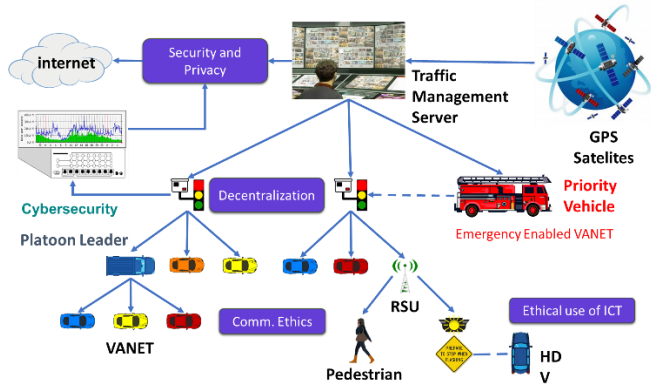

Fig. 10. Hierarchical-Ethical Network Layer Architecture

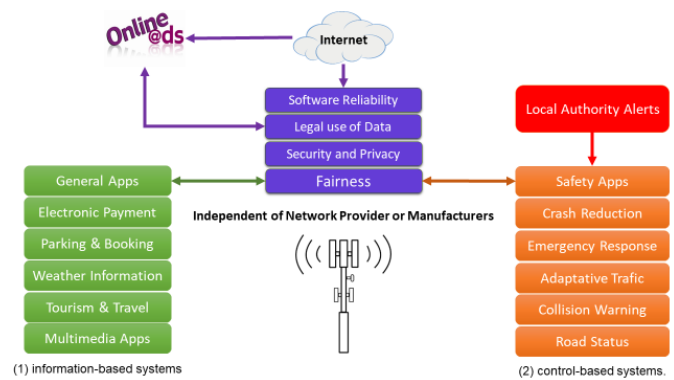

Fig. 11. Ethical Application Layer Architecture

## V. Proposed Computational Implementation

As previously stated [27], a SV ethical engine should be programed such that when confronted with a trolley like problem, the system will attempt to maximize the greater good. It has already been stated that this is a Sensing Layer problem that cannot be subordinated to external intervention. The question is, how do we define the greater good in an ethical calculation? DALY was suggested as a potential numerical indicator for utilitarian calculation and decision support [28]. Traditional DALY is computed as a function of life expectancy, this is interesting for epidemiological purpose but might not be optimal for decision support. Besides, DALY does not take into account vehicular safety ratings, or the relative forces involved in a collision. A specific quantitative measure for utilitarian calculations needs to be devised for SIoV.

NHTSA has established the 5-Star Safety Ratings system, to determine how well vehicles protect drivers and passengers during a crash [46]. A direct numerical conversion can be made and a 5 point vehicle considered the safest. Motorcycles and Bicycles could be considered to provide a safety ratings of 0.5 to 1.5 and a pedestrian could be considered to have a rating of 0.1. Other ratings would need to be determined for strollers, wheelchairs, skateboards, etc.

Using a Likert like scale, DALY can be simplified into 5 categorical segments: Infancy (5) – Childhood (4) – Adolescence & Early Adulthood (3) – Adulthood (2) – Mature Adulthood (1). Dividing Categorical DALY by the Vehicle Safety Rating, a Personal Ethical Value (PEV) is obtained, and expressed in utilitarian units (u). PEV for an infant in the safest vehicle will be 1u, whereas PEV for a Mature Adult in the same vehicle will be $\frac{1}{5}$u. The Total Ethical Value (TEV) for a vehicle can be calculated as the sum of all PEV in the vehicle ($TEV = \sum PEV_i$), the greater the TEV, the greater the ethical value associated to that entity. For example a pedestrian child will have a PEV of 40u and a 2 star car with two young adults and a child will have a TEV of 5u.

TEV accounts for part of the problem, since it allows to compute the greater good in a utilitarian calculation; however TEV is not enough, since potential damage is a function of effective crash force, and this in turn is a function of mass, speed, and breaking distance. The greater the force, the greater the potential damage. In case of utilitarian calculations, estimated crash force of each mobile should be multiplied by TEV, and the problem of computing the greater good becomes a problem of utilitarian force (UF) minimization in a free body diagram.

Fig. 4 provides an evaluation of a UF equilibrium. Assume that the car is 1m apart from all the potential targets and that the crash force at 1m would be 500N. Let's also assume that the truck is driven by an adult, the motorcycle by a young adult and both have a safety rating of 1; the car with two young adults and a child has a safety rating of 2. In a collision the total utilitarian force TUF is the addition of the absolute value of the UFs of the elements involved in the collision, as shown in fig. 12.

From the diagram the minimum damage (1,500uN) is produced by a collision between the car (V) and the Truck (B), while the maximum damage (42,500uN) is a direct collision between the car (V) and the children (A). The result is consistent with the expectations obtained by utilitarian analysis of fig. 4, however in this case there is a normalized utilitarian physical calculation that can be replicated and reviewed in a court of law.

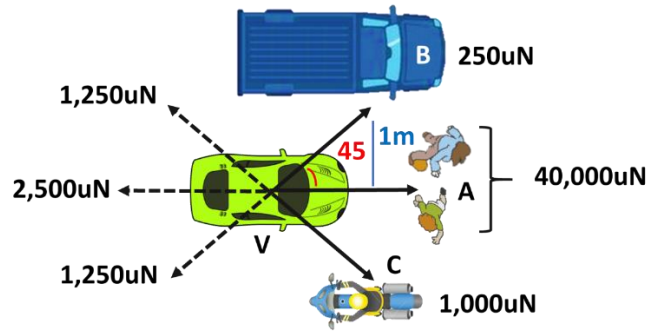

Fig. 12. Utilitarian Force (UF) equilibrium diagram for a car (V) with a safety rating of 2, occupied by two young adults and a child in potential collision trajectory against two pedestrian children, a truck or a motorcycle. The car is 1m apart from all the potential targets and the crash force at 1m is 500N. The truck is driven by an Adult, the motorcycle by a young adult and both have a safety rating of 1.

## Conclusion

In principle, SIoVs have the potential to improve vehicle and road safety, traffic efficiency and driver and passenger comfort. There is no margin of error for safety-critical technologies, and consumers will not give up control until they are confident that their vehicles and the mobile environment are safe and reliable at the same time. However, 100% efficacy is an impossible scenario, nonetheless automated road vehicles could predict various crash course alternatives and choose a path with the least human damage or probability of collision. The lowest damage in most cases includes damage to the vehicle and passengers rather than damage to pedestrians. The reluctance of passengers to implement ethical engines beneficial to pedestrians could delay the adoption of SIoV-compatible technologies. Several ethical concerns have been highlighted in this article based on traditional layered architecture. Each layer of the SIoV architecture and the entities associated with it are carefully examined to discuss their role and related ethical implications. Finally, a SIoV Ethical Architecture is proposed which takes into account the major ethical rules to be implemented before the development and deployment of SIoV systems.

The proposed computational implementation introduces the concept of utilitarian units, the advantage of including a physical unit is that it can be used in equations, which can be operationalized in an Ethical Engine, and analyzed in forensic investigations. Ethical Value could also be assigned to vehicles, animals, properties and other physical entities and used to optimize collision trajectories if needed, however, in a fully competent SIoV, once a vehicle has established a collision path, other vehicles would be automatically broadcasted and they can compute alternative paths to minimize the overall damage! The work presented in this article should lay the foundation for the development of ethical models for smart vehicles, RSUs, sensors and the overall ITS architecture.

The question is, if a passenger would accept an ethical engine where that utilizes a mathematical utilitarian engine, or would a passenger select a vehicle with another type of ethical engine? If multiple ethical engines were available for a particular vehicle, what type of engine would the owner choose? For example, participants interviewed in a group of six studies, concluded that "they would prefer to ride in SVs that protect their passengers at all costs; and would disapprove of enforcing SVs regulations that sacrifice their passengers for the greater good" [5].

If human beings are going to disapprove of ethical outcomes, should the ethical engine be defined by a regulating body as part of the SIoV operational protocol? According to [47] that might very well be the case, and they present the following conundrum:

- If driverless cars aren't safer than human drivers it will be unethical to sell them.
- Once driverless cars are safer than human drivers (reduce the risks to 3rd parties), driving will be unethical.

By following their line of thought and minimizing the death toll in case of unavoidable circumstances, once SVs are safer than human drivers, citizens will increase their willingness to accept self-sacrifice as being legally enforced. So, this ethical dispute can produce political pressure to be reflected in the legislation [46]. Overall, a SIoV architecture will be equipped with a norm baseline before being deployed, but this will not suffice for extended periods of time. The system must be capable of updating baseline system, because laws and regulations are going to change over time. In fact, we are not aware of the impact that smart infrastructure will have on the behavior of human beings. Counting with utilitarian units and utilitarian logic should simplify this task. As smart vehicles become smarter other ethical dilemmas will certainly develop, so smart infrastructure needs to be able to learn and regulate in response to social interactions.

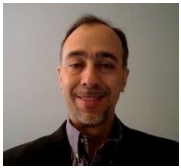

**Ricardo Silva** Member (M) of IEEE in 2001, and Senior Member (SM) in 2015. Dr. Silva was born in Caracas, Venezuela; earned his BS in Electronic Engineering and MSc in Biomedical Engineering, both from Universidad Simon Bolivar in Venezuela, and his PhD in Integrative Biosciences from Penn State University in USA.

He is VP for Innovation at the Foundation for Living, Wellness, and Health in USA and CSO of the Montenegro Institute for Cognitive Disabilities in Ecuador. He was president of July 17 Technical College at Yachay, the City of Knowledge, and Senior Researcher for the Prometheus Program, sponsored by the Secretariat for Higher Education, Science, Technology, and Innovation (SENESCYT), in Ecuador. He was Department Chair for the Department of Biological Process, Biophysics and Bioengineering Section at Simon Bolivar University in Venezuela. Lead the Health Care Technology Group (UGTS) and the Integrative Biosciences Laboratory (IBL) both at Simon Bolivar University. He has authored many works in clinical and biomedical engineering, health information technology and neuroscience amongst other areas.

Dr. Silva is also member of IFMBE, HIMSS, ACCE and AAMI. He has held various offices within IEEE both in Venezuela and Ecuador, including president of the EMBS chapter in both countries.

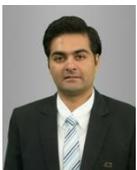

**Razi Iqbal** is an Associate Professor at the College of Computer Information Technology at the American University in the Emirates. Dr. Razi earned his PhD and Master's degree in Computer Science and Engineering from Akita University in Akita, Japan. He has published several papers in the areas of Computer Science and Engineering, specifically in computer networks, wireless technologies and wearable computing. His current research areas are use of short range wireless technologies in Precision Agriculture, Transportation and Education. Prior to working at AUE, he served as Chairman of the Department of Computer Science and IT at Lahore Leads University, Lahore, Pakistan. He has also served as Director ORIC (Office of Research, Innovation and Commercialization) at the same institute. During his tenure at Lahore Leads University his responsibilities were to administer the Computer Science and IT Department, design research policies for the university and teaching at undergraduate and graduate levels. He has also worked as a Research Scientist at Al-Khawarizmi Institute of Computer Science (KICS), University of Engineering and Technology (UET), Lahore, Pakistan.

Dr. Razi is serving as a reviewer to several peer reviewed journals and an EX-COM member to prestigious International Conference on Open Source Systems and Technologies held every year under the umbrella of IEEE. He is currently a member of IEEE and IEEE computer and computational society.